\documentclass[a4paper, 11pt, fleqn]{article}
\usepackage{epsfig}
\usepackage{amssymb}
\usepackage{amsmath}
\usepackage{graphicx}
\begin{document}

\Large
\begin{center}
\textbf{MULTIPLICATIVE STOCHASTIC MODEL OF THE TIME INTERVAL
BETWEEN TRADES IN FINANCIAL MARKETS }\\[1.25\baselineskip]
\small \textbf {V. Gontis}

\vskip 2mm Institute of Theoretical Physics and Astronomy \\ A.Gostauto 12, 2600 Vilnius, Lithuania \\
gontis@ktl.mii.lt
\end{center}
\small

\normalsize

\begin{abstract}
Stock price change in financial market occurs through transactions
in analogy with diffusion in stochastic physical systems. The
analysis of price changes in real markets shows that long-range
correlations of price fluctuations largely depend on the number of
transactions. We introduce the multiplicative stochastic model of
time interval between trades and analyze spectral density and
correlations of the number of transactions. The model reproduces
spectral properties of the real markets and explains the mechanism
of power law distribution of trading activity. Our study provides
an evidence that statistical properties of financial markets are
enclosed in the statistics of the time interval between trades.
Multiplicative stochastic diffusion may serve as a consistent
model for this statistics.

\end{abstract}

\section{Introduction}

Complex collective phenomena usually are responsible for
power-laws which are universal and independent of the microscopic
details of the phenomenon. Examples in physics are numerous.
Power-laws are intrinsic features of the economic and financial
data as well. The aim of this contribution is to analyze a
relation between the origin of the power law distribution and the
power-law correlations in financial time series. There are
numerous studies of power-law probability distributions in various
economic systems \cite{1,2,3,4}.
 The time-correlations in financial time series are studied extensively as well
\cite{5,6,7}. The random multiplicative process built into the
model of wealth distribution yields Pareto style power law
\cite{2}. The generalized Lotka Volterra dynamics developed by
S.Solomon and P.Richmond is in the use for various systems
including financial markets \cite{1} . However, these models
generically lead to the non-universal exponents and do not explain
the power-law correlations in financial time series \cite{8}.
Recently we adopted the model of $1/f$ noise based on the Brownian
motion of time interval between subsequent pulses, proposed by
Kaulakys and Meskauskas \cite{9,10,11,12} to model share volume
traded in financial markets \cite{13} . The idea to transfer long
time correlations into stochastic process of the time interval
between trades or time series of trading activity is in
consistence with the detailed studies of the empirical financial
data \cite{5,6} and fruitfuly reproduces spectral properties of
financial time series \cite{13}. However, the investigation of the
model revealed,  that the simple additive Brownian model of time
interval between trades failed to reproduce power-law probability
density distribution (pdf) of trading activity. On the other hand,
several authors showed empirically that the fluctuations of
various financial time series possess multifractal statistics
\cite{14,15,16} . Therefore we introduce the stochastic
multiplicative model of time interval between trades and analyze
the statistical properties of the model trading activity
numerically and analytically. We show the consistence of the
approach with the results of statistical analyzes of the empirical
financial time series. This sheds light on the relation between
the power-law probability distribution and the power-law
correlations in financial time series.

\section{Multiplicative time interval model}

Following our previous model \cite{13} we consider a signal $%
I(t)$ as a sequence of the random correlated pulses
\begin{equation}
I(t)=\sum_{k}q_{k}\delta (t-t_{k})
\end{equation}
where $q_{k}$ is a contribution to the signal of one  pulse at the
time moment $t_{k}$, for example, a contribution of one
transaction to financial data. Signal (1) was introduced by
Kaulakys and Meskauskas for the modelling of $1/f$ noise and could
be used in a large variety of systems with the flow of point
objects or subsequent actions. In the statistical data analysis we
usually deal with the discrete time series in equal time intervals
$\tau _{d}$. Integrating the signal $I(t)$ in the subsequent time
intervals of length $\tau _{d}$ we get discrete time series and,
in analogy with financial time series, we call it the volume
$V_{j}$
\begin{equation}
V_{j}=\int\limits_{t_{j}}^{t_{j}+\tau
_{d}}I(t)dt=\sum\limits_{t_{j}<t_{k}<t_{j}+\tau _{d}}q_{k},\ \ \ \ \ \ \
t_{j}=j\tau _{d}.
\end{equation}
We can define the number of trades $N_{j}$ in the time interval
$t_{j}\div
(t_{j}+\tau _{d})$ by the same formula (2), when $q_{k}\equiv 1$ , $%
N_{j}\equiv V_{j}$ . The power spectral density $S(f_{s})$ of the signal $%
V_{j}$ is defined as
\begin{equation}
S(f_{s})=\left\langle \frac{2}{\tau _{d}n}\left|
\sum\limits_{j=1}^{n}V_{j}\exp \left\{ -i2\pi (s-1)(j-1)/n\right\}
\right| ^{2}\right\rangle
\end{equation}
where $f_{s}=(s-1)/T$ , $T=\tau _{d}n$.
 In this paper we
investigate statistical properties of the time series $N_{j}$,
when the sequence of event time $t_{k}$ is generated by the
multiplicative stochastic process. There are some arguments for
this choice of interevent time model. First of all the
multiplicativity is an essential feature of the processes in
economics \cite {1,2,3,4} . Multiplicative stochastic processes
yield multifractal intermittency and are able to produce power-law
pdf. Pure multiplicative processes are not stationary and
artificial diffusion restriction mechanisms have to be introduced.
Let us start our analysis introducing the multiplicative
process with the logaritmic diffusion restriction for the large deviations of $%
\tau _{k}$ from\ the mean value $\overline{\tau }_{k}=1$
\begin{eqnarray}
t_{k+1} &=&t_{k}+\tau _{k+1}  \nonumber \\
\tau _{k+1} &=&\tau _{k}(1-\gamma \ln \tau _{k}+\sigma \varepsilon _{k}).
\end{eqnarray}
Here the interevent time $\tau _{k}$ fluctuates due to the
external random perturbation by a sequence of uncorrelated
normally distributed random variables $\left\{ \varepsilon
_{k}\right\} $ with zero expectation and unit variance, where
$\sigma $ denotes the standard deviation of the white noise and
$\gamma <<1$ is a damping constant. The choice to introduce
multiplicative logaritmic diffusion damping $\gamma \tau _{k}\ln
\tau _{k}$ retains symmetry in logaritmic scale and multiplicative
nature of the process for small values of $\gamma \ln \tau _{k}$ .
Long time stationary pdf of $\tau _{k}$ is a Lognormal
distribution
\begin{equation}
P(\tau )=\frac{2}{\sqrt{2\pi \sigma ^{2}/\gamma }}\frac{1}{\tau
}\exp \left\{ -\frac{2\gamma }{\sigma ^{2}}\ln ^{2}\tau \right\}.
\end{equation}
We derive the distribution (5) from the solution of the
generalized Langevine equation \cite{17}
\begin{eqnarray}
\stackrel{.}{x} &=&F(x)+G(x)\eta (t), \ \ -\frac{\partial V}{\partial x}=%
\frac{F}{G^{2}},  \nonumber \\
p(x) &=&C\exp \left[ -\left\{ \frac{V(x)}{\frac{1}{2}\sigma
^{2}}+\ln G(x)\right\} \right]
\end{eqnarray}
where $\left\langle \eta (t)\right\rangle =0$ \ and $\left\langle
\eta (t)\eta (t^{\prime})\right\rangle =\sigma ^{2}\delta
(t-t^{\prime})$. Note that Lognormal distribution is invariant to
the change of variables: $n=1/\tau $ and in the wide range of
argument values $\left| \ln \tau \right| <<\sigma
/\gamma ^{1/2}$ is equivalent to the power-law $\sim%
1/\tau $. Nevertheless, we have to point out that lognormal distribution of $%
\tau _{k}$ is not appropriate to reproduce the power-law
distributions of variables in financial markets. We expect that
iterative stochastic model (4) producing universal lognormal
distribution\ will be useful for other applications. We base our
model on the generic multiplicative process
\begin{equation}
\tau _{k+1}=\tau _{k}+\gamma \tau _{k}^{2\mu -1}+\tau _{k}^{\mu }\sigma
\varepsilon _{k},
\end{equation}
as the most appropriate seeking to reproduce power-law probability
density distributions of variables in financial markets. In this
paper we will use a very simple $\tau _{k}$ diffusion damping
model assuming that $0\leq \tau _{k}\leq 1$ and $0<\gamma <<\sigma
$. More sophisticated damping models can be introduced in further
developments of this approach. The most natural choice is to
assume $\mu =1$, giving pure multiplicativity of $\tau _{k}$. We
decided to keep this parameter in our notation seeking to sustain
generic definition. In the $k$ scale  Eq. (7) defines generalized
Langevine equation (6) with $F(\tau )=\gamma \tau _{k}^{2\mu -1}$
and $G(\tau )=\tau _{k}^{\mu }$. Note that the solution (6) of the
generalized Langevine equation is based on the Stratanovich
convention. Our numerical analysis confirms the preference of the
Stratanovich convention in comparison with Ito convention
regarding discrete iterative equation (7).  The stationary pdf of
$\tau _{k}$ defined by iterative relation (7) can be derived from
the solution (6) of the generalized Langevine equation,
\begin{equation}
P_{k}(\tau )=\frac{C_{\tau}}{\tau ^{\mu -2\gamma /\sigma ^{2}}}
\end{equation}
where the normalization constant $C_(\tau)$ can be defined from the integral $\int\limits_{0}^{1}P_{k}(%
\tau )d\tau =1$. The stationary distribution function for the number of trades $%
N_{k}$ in the time interval $\tau _{d}$ can be derived from
(8). Simple change of variables $N=\tau _{d}/\tau $ in (8) having in mind that $%
P_{k}(\tau )d\tau =$ $P_{k}(N)dN\ \ $, gives probability
distribution function of N
\begin{equation}
P_{k}(N)=\frac{C_{N}}{N ^{2-\mu+2\gamma /\sigma ^{2}}}
\end{equation}
where the time interval  $\tau _{d}$ is included in normalization constant $%
C_{N}$. Note that variance $\langle N_{k}^{2}\rangle-\langle
N_{k}\rangle^{2}$ diverges when $2\gamma/\sigma^{2}<\mu+1$. We
will see later that the transition from the stationary regime to
the non-stationary one is essential for the appearance of $1/f$ \
power spectral density. Distribution function (9) can be
transformed to the time scale taking into account that every step
in the iteration equation (7) is equal to the time interval $\tau
_{k}$. In the real time scale the pdf of $N$ takes the form
\begin{equation}
P_{t}(N)=\frac{C_{N_{t}}}{ N^{3-\mu+2\gamma /\sigma^{2}}}
\end{equation}
Here $C_{N_{t}}$ is a new normalization constant.

\section{Numerical results and discussion}

\vspace{1pt}We have already introduced the multiplicative
stochastic process as a model for the time interval between trades
in financial markets. This yields the power-law distribution for
the number of trades per constant time interval $\tau _{d}$,
equation (10). The exponent of cumulative power-law distribution
$\alpha$ for pure  multiplicative model, $\mu =1$, depends only on
the parameter $\nu =2\gamma /\sigma ^{2}$, i.e.,
\begin{equation}
\alpha =1+2\gamma /\sigma ^{2}.
\end{equation}
Parameter $\nu$ defines the
ratio of damping constant $\gamma$ to the coefficient of stochastic diffusion $\frac{1%
}{2}\sigma ^{2}$. In this chapter we will analyze spectral
properties of the model and related autocorrelation of trading
activity $N$. The model presented can be easily calculated
numerically using equations (1), (2), (3) and (7). We smooth the
power spectral density with standard moving average procedure. We
present the numerically calculated power spectral density  $S(f)$
of $N$ for $\nu =2.4;2.0$ and $1.5$  on the Fig. 1. (a), (b) and
(c) respectively with $\sigma =0.1$ and $\tau _{d}=10$. For the
values of
parameter $\nu >2$, we observe the power spectral density  $S(f)%
\sim1/f^{1.5}$. For $\nu \simeq 2$ we obtain $S(f)\sim1/f$ and we
get the slope of $S(f)\sim1/f^{0.5}$ when $\nu <2$.
\begin{center}
\begin{figure}[tbp]
  \begin{center}
  \includegraphics [width=1.0\hsize] {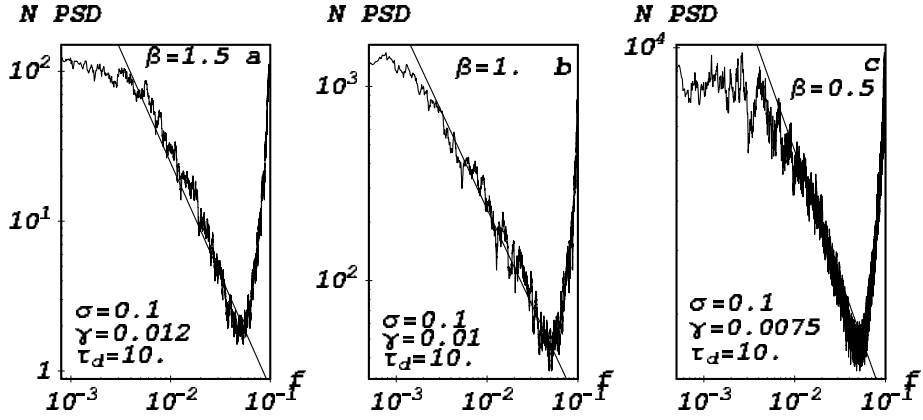}
  \end{center}
  \caption{Power spectral density  of trading activity $N$ (N PSD) versus
frequency $f$ calculated from the model described by Eqs. (1),
(2), (3) and (7) with $\mu =1$. The sinuous curves represent
results of the
numerical simulations smoothed with standard moving average procedure over 43 points with $%
\sigma =0.1$ and $\tau _{d}=10$, while the straight lines
represent power-law power spectral density $S(f)\sim1/f^{\beta }$
with the various slopes $\beta $. For $\gamma =0.012$, $\beta
=1.5$ (a); for $\gamma =0.01$, $\beta =1$ (b) and  for $\gamma
=0.0075$, $\beta =0.5$ (c).}
  \label{Picture1}
\end{figure}
\end{center}
Numerical results confirm that $1/f$ noise occurs when stochastic
process defining the trading activity $N_{k}$ experience the
transfer from the stationary to the non-stationary regime. This
general rule is probably applicable to the various models of the
time intervals between pulses and is in agreement with the theory
of self-affine fractals, see \cite{18}. It should be noted that
transition from slope $\beta =1.5$ to $\beta =0.5$ of the power
spectral density is very sharp with the change of parameters $\nu
$ or $\gamma $ less then twice. Calculations with different values
of parameters confirm a universal nature of the relation between
the slope of the power spectral density $\beta $ and parameter of
multiplicative diffusion $\nu $ . This relation is independent of
$\tau _{d}$ in wide range of values. We present our numerical
results of power spectral density for different sets of parameters
in Fig. 2. The slopes $\beta =1.5;1.0 and 0.5$, of power spectral
density are related with the corresponding values of parameter
$\nu =2.4;2.0 and 1.6$, as in  Fig. 1. Numerous calculations
confirm that the slope of power spectral density $\beta $ is
defined by $\nu $, exhibiting sharp transition from $\beta =1.5$
to $\beta =0.5$, when $\nu \simeq 2.0$. This transition is related
with divergence of $N_{k}$variance $\langle
N_{k}^{2}\rangle-\langle N_{k}\rangle^{2}$ defined from (9).
\begin{center}
\begin{figure}[tbp]
  \includegraphics [width=1.0\hsize] {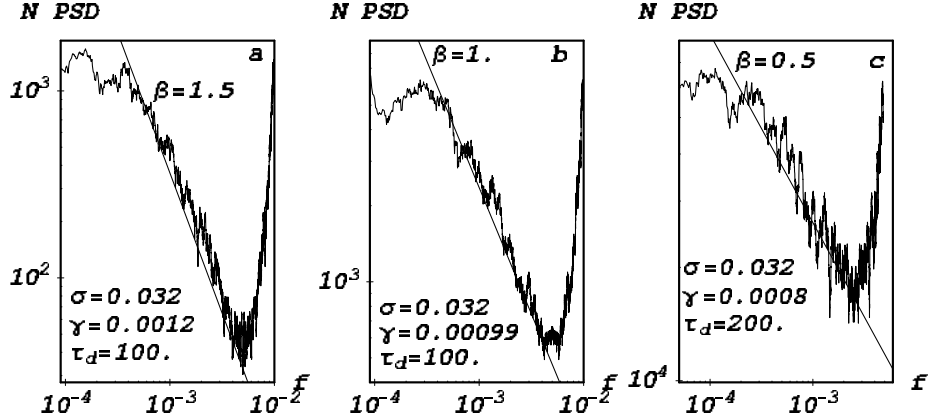}
  \caption{The same as in fig. 1 but with parameters  $%
\sigma =0.032$ and $\tau _{d}=100$ or $200$. The straight lines
represent power-law power spectral density $S(f)\sim1/f^{\beta }$
with the
various slopes $\beta $. For $\gamma =0.0012$, $\beta =1.5$ (a); for $%
\gamma =0.001$, $\beta =1$ (b); for $\gamma =0.0008$, $\beta =0.5
$ (c).}
  \label{Picture2}
\end{figure}
\end{center}
In the previous section we derived the power-law probability
distribution function of $N$, Eq. (10). Let us compare the
exponent of cumulative distribution from (10), $\alpha =1+\nu $,
with numerical results in the interval of $\nu $ values $1.5\div
3.5$ related with the slope $\beta $ change. In Fig. 3. we present
numerically calculated $\alpha $ as a function of $\nu $ for
various values of $\sigma ,\gamma $, and $\tau _{d}$. The least
square linear approximation of numerical results, straight line,
yields that $\alpha =0.96+0.97\nu $. It shows that the probability
distribution function (10) derived from the generalized Langevine
equation (6)  describes  the power-law distribution of $N$ very
well and serves as an evidence that the Stratanovich convention is
appropriate for the multiplicative stochastic model (7). For $\nu
\simeq 2$, corresponding to $\beta =1.$,  we get the value of
$\alpha \simeq 3.$ This is in good agreement with empirical data
of power-law distribution of trading activity $N$ in the financial
markets $\alpha \simeq 3.4$ \cite {5}.
\begin{center}
\begin{figure}[tbp]
\begin{center}
  \includegraphics [width=0.6\hsize] {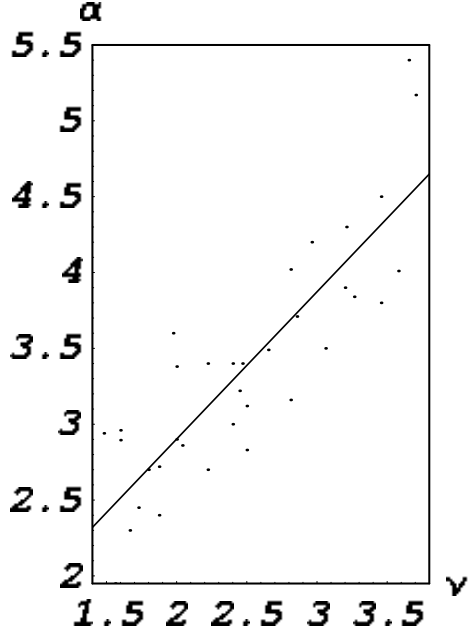}
  \end{center}
  \caption{The exponent $\alpha $ of power-law distribution
  function for $N$
 versus multiplicative diffusion parameter $\nu $
calculated numerically from the model described by Eqs. (1), (2),
(3) and (7). Points
represent various realizations with the parameters $\sigma =0.05\div 0.1$, $%
\gamma =0.001\div 0.015$, $\tau _{d}=10\div 100$. Straight line
represents
the least square approximation to the numerical data $\alpha =0.96+0.97\nu $%
.}
  \label{Picture3}
\end{figure}
\end{center}
Numerical results confirm that multiplicative stochastic model of
the time intervals between trades in the financial markets is able
to reproduce the main statistical properties of trading activity
$N.$ First of all, even in a very simple multiplicative diffusion
restriction $0\leq \tau _{k}\leq 1$, the model reproduces the
power-law distribution function with $\alpha \simeq 3$. The
power-law exponent is related with the parameter $\nu =2\gamma
/\sigma ^{2}$, which defines the distribution functions of time
interval between trades $\tau $, and of trading activity $N$ as
well as the slope $\beta $ of its' power spectral density
$S(f)\sim1/f^{\beta }$. This sheds light on the origin of
universal nature of the power-law distributions of such variables
as trading activity in financial markets $N$ and its' power
spectral density $S(f)$.

\section{Conclusions}

We have introduced multiplicative stochastic model of the time
interval $\tau $ between trades in the financial markets Eq. (7)
with the diffusion restriction in the interval $0\leq \tau
_{k}\leq 1$. This defines stochastic fluctuations of trading
activity $N$, Eq.
(2), and reproduces its' power-law distribution of probability and the slope of power spectral density $S(f)%
\sim1/f^{\beta }$. The comparison of the empirical data from the
financial market analysis with the model proposed implies that
markets function in the area of parameters, where convergence of
$N_{k}$ variance disappears and $1/f$ noise of $N$ occurs. This
rather a sharp transition to the non-stationary regime of $ N_{k}$
fluctuations probably is the indispensable feature of the
financial markets. This relates the exponent of the power-law
distribution to the slope $\beta $ of power spectral density into
universal interdependence. It implies that the main statistical
properties of the financial markets have to be defined by the
fluctuations of time interval between the trades accumulating the
power-laws and the long time correlations. Further empirical
analysis of $\tau $ statistics and the adjusted specification of
the model is desirable. We do expect that multiplicative model of
time interval between the trades with more specific diffusion
restriction conditions and more precisely adjusted $\mu$ lies in
the background of the financial market statistics and can be very
useful in financial time series analysis.

\section{Acknowledgment}

The author gratefully acknowledges the numerous and long
discussions with Bronislovas Kaulakys regarding inherent origin of
$1/f$ noise and various interpretations of stochastic differential
equations.



\begin{thebibliography}{99}
\bibitem{1}  Blank A., Solomon S., Power-laws in cities population,
financial markets and internet sites (scaling in systems with variable
number of components), Physica A, 287 (1-2), (2000), p. 279-288.

\bibitem{2}  Levy M., Solomon S., Power-laws are logaritmic Boltzman laws,
Mod.Phys. C, \textbf{7, }4, (1996), p. 595.

\bibitem{3}  M. Levy M, and S. Solomon (1997), Physica A 242, 90

\bibitem{4}  Sornette D., Cont R., J. Phys. I (France) \textbf{7} (1997), p.
431

\bibitem{5}  Gopikrishnan P., Plerou V., Liu Y., Amaral L.A.N., Gabaix X.,
Stanley H.E., Scaling and correlation in financial time series, Physica A
\textbf{287, }(2000), p. 362-373.

\bibitem{6}  P. Gopikrishnan et al.\textbf{, }Phys. Rev. E, V. 62 (4),
(2000), p. R4493-R4496.

\bibitem{7}  Cizeau P., Potters M., Bouchaud J.\textbf{,} Science,
Correlation structure of extreme stock returns, Quantitative Finance \textbf{%
1}, (2001), p. 217-222.

\bibitem{8}  Bouchaud J.-P. Power-laws in economy and finance: some ideas
from physics, e-print: cond-mat/0008103 v.1.

\bibitem{9}  B. Kaulakys and T. Me\v{s}kauskas,\textbf{\ }Phys. Rev. E, V
58, p.7013 (1998).

\bibitem{10}  B. Kaulakys and T. Me\v{s}kauskas\textbf{, }Nonlinear
Analysis: Modelling and Control, Vilnius, IMI, 1999, No 4.

\bibitem{11}  B. Kaulakys, Autoregressive model of 1/f noise, Phys.
Lett. A, \textbf{257, }(1999), p. 37-42.

\bibitem{12}  B. Kaulakys, On the intrinsic origin of 1/f noise,
Microelectronics Reliability, \textbf{40}, (2000), p. 1787-1790.

\bibitem{13}  Gontis V., Modelling share volume traded in financial
markets,Lithuanian J. Phys.,\textbf{41}, (2001), p. 551-555,
cond-mat/0201514 (2002).

\bibitem{14}  Schmitt F., Schertzer D., Lovejoy S., Multifractal analysis of
foreign exchange data, Appl. Stoch. Mod. Data Anal. \textbf{15},
(1999), p. 29-53.

\bibitem{15}  Schmitt F., Schertzet D., Lovejoy S., Multifractal
fluctuations in finance, Int. J. Thoer. Appl. Fin., \textbf{3}, No
3, (2000), p. 361-364.

\bibitem{16}  Vandewalle N., Ausloos M., Multi-affine analysis of typical
currency exchange rates, Eur. Phys. J., \textbf{B4}, (1998) p. 257-261.

\bibitem{17}  Richmond P., Solomon S., Power-laws are Boltzman Laws in
Disguise, e-print: cond-mat/0010222 (2000).

\bibitem{18}  Malamud B.D., Turcotte D.L., Self-affine time series: measures
of weak and strong persistence, Stat. Planing and Inference,
\textbf{80}, (1999), p. 173-196.
\end{thebibliography}
\end{document}